\documentclass[preprint,aps,tightenlines]{revtex4}
\usepackage[dvipdf]{graphicx}

\begin{document}
\title{Lithium atom interferometer using laser diffraction: description and experiments}

\author{A. Miffre, M. Jacquey, M. B\"uchner, G. Tr\'enec and J. Vigu\'e}
\address{ Laboratoire Collisions Agr\'egats R\'eactivit\'e -IRSAMC
\\Universit\'e Paul Sabatier and CNRS UMR 5589
\\ 118, Route de Narbonne 31062 Toulouse Cedex, France
\\ e-mail:~{\tt jacques.vigue@irsamc.ups-tlse.fr}}

\date{\today}

\begin{abstract}

We have built and operated an atom interferometer of the
Mach-Zehnder type. The atomic wave is a supersonic beam of lithium
seeded in argon and the mirrors and beam-splitters for the atomic
wave are based on elastic Bragg diffraction on laser standing
waves at $\lambda=671$ nm. We give here a detailed description of
our experimental setup and of the procedures used to align its
components. We then present experimental signals, exhibiting
atomic interference effects with a very high visibility, up to
$84.5\pm 1$ \%. We describe a series of experiments testing the
sensitivity of the fringe visibility to the main alignment defects
and to the magnetic field gradient.


\end{abstract}

\maketitle


\section{Introduction and brief historical overview}

We have built a Mach-Zehnder atom interferometer, which gave its
first signals in 2001 \cite{delhuille02a}. In this interferometer,
the atomic wave is a supersonic beam of lithium seeded in argon,
with a lithium de Broglie wavelength $\lambda_{dB} = 54$ pm.
Coherent atom manipulation is based on Bragg diffraction on
quasi-resonant laser standing waves at a wavelength $\lambda_L
\approx 671$ nm. We use elastic laser diffraction, which can be
made with ordinary single frequency lasers, because this process
has little sensitivity to the phase noise of the laser beams.
However, the associated difficulty is that the output atomic beams
differ only by their directions of propagation and not by their
internal states. Therefore, such an interferometer must be
operated with a highly collimated atomic beam resulting in a
strongly reduced output atomic flux. Fortunately, the transmission
of such a Bragg Mach-Zehnder interferometer is quite high and,
thanks to an intense lithium beam and a very sensitive hot-wire
atom detector, we obtain reasonably large signals. Moreover, we
have been able to observe interference signals while using the
diffraction orders $p=1$, $2$ and $3$ and in the case of the first
order, the signal exhibits an excellent fringe visibility
${\mathcal{V}} = 84.5\pm 1$ \%.

We may recall the development of atom interferometry since 1991,
when several atom interferometers gave their first signals:
\begin{itemize}
\item a Young's double slit experiment by O. Carnal and J. Mlynek,
with a supersonic beam of metastable helium \cite{carnal91}

\item a Mach-Zehnder interferometer by D. Pritchard and co-workers
using a thermal atomic beam of sodium and diffraction on material
gratings \cite{keith91}

\item  a Ramsey-Bord\'e interferometer by J. Helmcke and
co-workers, with a thermal atomic beam of calcium, was used to
demonstrate Sagnac effect with atomic waves \cite{riehle91}

\item an atom interferometer using Raman diffraction by M.
Kasevich and S. Chu, with cold sodium atoms, was used to make the
first high sensitivity measurement of the local acceleration of
gravity by atom interferometry \cite{kasevich91}
\end{itemize}

This research field has been rapidly expanding since 1991 and an
excellent overview of this field and of its applications can be
found in the book "Atom interferometry" \cite{berman97} published
in 1997. Many types of atom interferometers have been developed
and we limit the present review to the apparatuses in which the
atomic paths are noticeably different, i.e. we will not discuss
the interferometers, such as atomic clocks, in which the momentum
transfer is very small. Moreover, we limit our review to
interferometers operating with thermal atoms or molecules, quoting
only the first publication for each interferometer. In addition to
the interferometers built in 1991, we find: a magnesium atom
interferometer by W. Ertmer and co-workers \cite{sterr92}; a
calcium atom interferometer by A. Morinaga and co-workers
\cite{morinaga93}; an I$_2$ molecular interferometer by Ch. J.
Bord\'e and co-workers \cite{borde94}; a Na$_2$ molecular
interferometer by D. Pritchard and co-workers \cite{chapman95}; a
metastable argon interferometer of A. Zeilinger and co-workers
\cite{rasel95}; a metastable neon interferometer by Siu Au Lee and
co-workers \cite{giltner95b}; a cesium atom interferometer
gyroscope by M. Kasevich and co-workers \cite{gustavson97}; a
$K_2$ molecular interferometer by E. Tiemann and co-workers
\cite{lisdat00}, a helium atom and dimer interferometer by J. P.
Toennies and co-workers \cite{toennies01}, a large molecule
interferometer by A. Zeilinger and co-workers \cite{brezger02}; a
metastable hydrogen interferometer by T.W. H\"ansch and co-workers
\cite{heupel02}.

In this paper, we recall the principles of Mach-Zehnder atom
interferometers and of laser diffraction. Then, we explain our
basic choices and we describe our setup and its alignment
procedures. We present a diffraction experiment, used to choose
the parameters of the laser standing waves, and a set of
interference signals recorded using the diffraction orders
$p=1,2,3$. We explain how we have optimized the fringe visibility
by a systematic study of its variations with the main defects of
the interferometer.

\section{Mach-Zehnder atom interferometers: general properties
and our design}

\subsection{General properties}

\begin{figure}[hbt]
  \begin{center}
  \includegraphics[width= 11.5 cm]{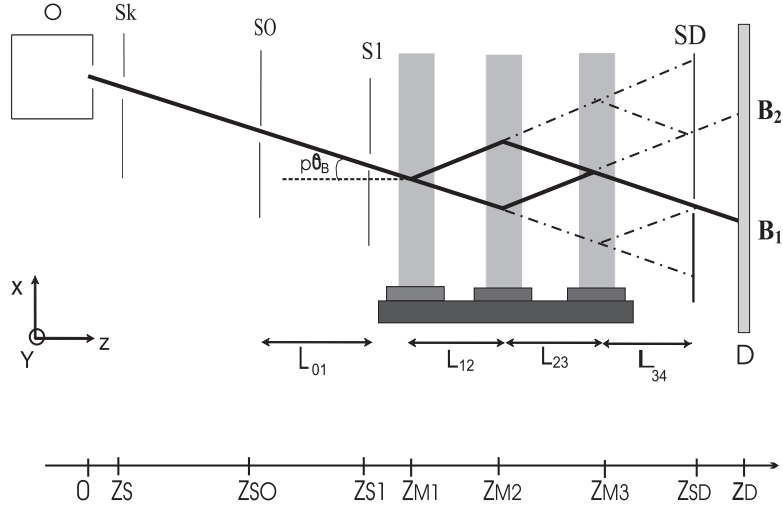}
 \caption{\label{MZschematic} Schematic drawing of our Mach-Zehnder
interferometer (top view). The $x,y,z$ axis are represented and we
give the distance $z$ of each element to the nozzle. O : lithium
oven; Sk : skimmer at $z_s= 20$ mm ; $S_0$: first collimating slit
at $z_{S0} = 485$ mm; $S_1$: second collimating slit at $z_{S1} =
1265$ mm; $M_i$, $i=1-3$ : mirrors for the laser standing waves at
$z_{M1} =1415$ mm, $z_{M2} = 2020$ mm and $z_{M3} = 2625 $ mm;
$B_1$ and $B_2$: complementary exit beams; $S_D$: detector slit
with a tunable width at $z_{SD} =3025$ mm ; $D$: $760$ $\mu$m wide
rhenium hot wire of the Langmuir-Taylor detector at $z_D= 3375$
mm. We have also represented the main stray atomic beams produced
by diffraction on the three gratings, assuming that only two
diffraction orders, $0$ and $p$ are produced, as in the ideal
Bragg regime.}
  \end{center}
  \end{figure}

A Mach-Zehnder grating interferometer is derived from the optical
Mach-Zehnder interferometer by replacing the beam-splitters and
mirrors by diffraction gratings. This interferometer was developed
with X-rays \cite{bonse65} in 1965, with neutrons \cite{rauch74}
in 1974 and with atoms \cite{keith91,kasevich91} in 1991. Figure
\ref{MZschematic} presents a schematic drawing of the atom paths
in such an interferometer. In the simplest approximation, the
incident atomic wave is treated as a plane wave $\Psi({\mathbf
r})= \exp\left[i{\mathbf k}\cdot{\mathbf r}\right]$ and
diffraction of order $p$ by grating $G_j$ produces also a plane
wave:

\begin{equation}
\label{g2} \Psi_d ({\mathbf r}) =  \alpha_j(p) \exp\left[i{\mathbf
k}\cdot{\mathbf r} + i p {\mathbf k}_{Gj}\cdot\left({\mathbf r}-
{\mathbf r_j}\right)\right]
\end{equation}

\noindent

Conservation of energy and momentum must be fulfilled and equation
(\ref{g2}) is exact only in the case of Bragg diffraction but,
near this geometry, it is valid up to the first order in power of
$k_{Gj}/k$. $\alpha_j(p)$ is the diffraction amplitude of order
$p$ by grating $G_j$. The wave vector $ {\mathbf k}_{Gj}$ of
grating $G_j$ lies in the grating plane, perpendicular to its
lines, with a modulus $ k_{Gj} = 2\pi/a$ ($a$ is the grating
period, assumed to be the same for the three gratings). ${\mathbf
r_j}$ measures the position of a reference point linked to grating
$G_j$. As shown by equation (\ref{g2}), the phase of the
diffracted beam depends rapidly on the position of the grating in
its plane. The output beam labeled $B_1$ in figure
\ref{MZschematic} results from the interference of two waves $
\Psi_u $ (following the upper path with the diffraction orders
$p$, $-p$ and $0$) and $\Psi_l $ (following the lower path with
the diffraction orders $0$, $p$ and $-p$):

\begin{equation}
\label{g3} \Psi_{u/l} ({\mathbf r})=  a_{u/l} \exp \left[i \left(
{\mathbf k}\cdot{\mathbf r}+ \varphi_{u/l}\right)\right]
\end{equation}

\noindent with $a_u= \alpha_1(p)\alpha_2(-p)\alpha_3(0)$ and $a_l
= \alpha_1(0)\alpha_2(p)\alpha_3(-p)$ while $\varphi_u =
p\left[{\mathbf k}_{G1}\cdot\left({\mathbf r}-{\mathbf r_1}\right)
-{\mathbf k}_{G2} \cdot\left( {\mathbf r} - {\mathbf
r_2}\right)\right]$ and $\varphi_l = p\left[{\mathbf
k}_{G2}\cdot\left({\mathbf r}-{\mathbf r_2}\right) -{\mathbf
k}_{G3} \cdot\left( {\mathbf r} - {\mathbf r_3}\right)\right]$.
These two waves interfere on the detector and the resulting total
intensity is given by integrating over the detector surface:

\begin{eqnarray}
\label{g5}
 I_1  = \int d^2 {\mathbf r}\left| \Psi_u + \Psi_l \right|^2
= \int d^2 {\mathbf r}\left[ a_u^2 + a_l^2 + 2 a_u a_l
\cos\left(\varphi_u - \varphi_l\right)\right]
\end{eqnarray}

\noindent To simplify the algebra, we have assumed that the
amplitudes $a_u$ and $a_l$ are real. The phase $\left(\varphi_u -
\varphi_l\right)$ is given by $\varphi_u - \varphi_l = p
\left[\Delta {\mathbf k}_G \cdot {\mathbf r} + \Delta
\varphi_G\right]$ where $\Delta  {\mathbf k}_G$ is given by:

\begin{equation}
\label{g61}\Delta {\mathbf k}_G = {\mathbf k}_{G1} + {\mathbf
k}_{G3} - 2{\mathbf k}_{G2}
\end{equation}

\noindent and the phase $\Delta \varphi_G$ is a function of the
grating positions only:

\begin{equation} \label{g7}
\Delta \varphi_G = \left[2 {\mathbf k}_{G2}\cdot {\mathbf r}_2
-{\mathbf k}_{G1} \cdot {\mathbf r}_1 - {\mathbf k}_{G3} \cdot
{\mathbf r_3}\right]\approx k_G \left(2 x_2 -x_1 -x_3 \right)
\end{equation}

\begin{figure}[hbt]
  \begin{center}
  \includegraphics[width= 11.5 cm]{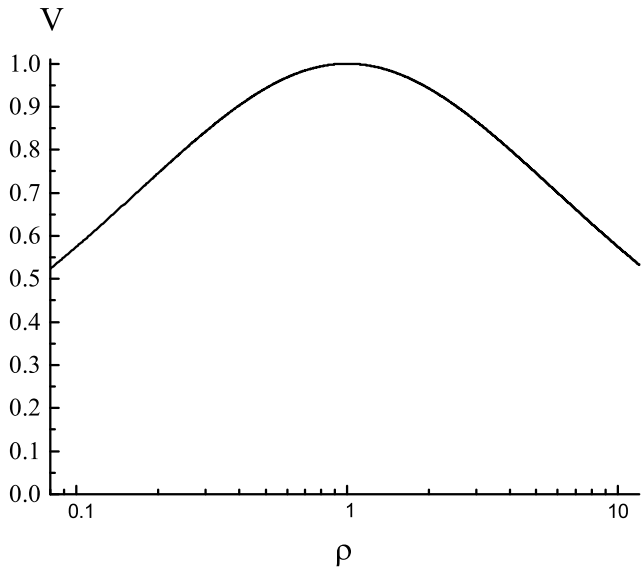}
\caption{\label{theorycontrast} Fringe visibility ${\mathcal{V}}$
for a two-beam interference as a function of the intensity ratio
$\rho$. A logarithmic scale has been used for $\rho$ so as to
exhibit the symmetry when $\rho$ is replaced by $1/\rho$.}
  \end{center}
  \end{figure}

\noindent Fringes appear over the detector area if the condition
$\Delta {\mathbf k}_G = {\mathbf 0}$ is not fulfilled. In the
experiments, this condition is verified by tuning the orientation
of one grating in its plane and any small deviation induces a
large visibility loss, as shown below (see figure \ref{deltak}).
The $x$-positions of the three mirrors change the phase $\Delta
\varphi_G$ of the atom interference fringes. This property
provides a very convenient method to sweep the interference
fringes: this phase is non-dispersive, i.e. independent of the
velocity of the atomic wave, so that there is no associated
visibility loss. If we assume that $\Delta {\mathbf k}_G =
{\mathbf 0}$, then $\left| \Psi_u + \Psi_l \right|$ is independent
of ${\mathbf r}$ and the intensity $I_1$ of the exit beam $B_1$ is
proportional to:

\begin{equation}
\label{g7}
 I_1 = a_u^2 + a_l^2 + 2 a_u a_l \cos( \varphi_u - \varphi_l) =
 I_{1,m} \left[1 + {\mathcal{V}}\cos( \varphi_u -
 \varphi_l)\right]
\end{equation}

\noindent where ${\mathcal{V}}$ is the fringe visibility given by:
\begin{equation}
\label{g8} {\mathcal{V}} = \frac{2a_u a_l}{a_u^2 + a_l^2} =
\frac{2\sqrt{\rho}}{1 + \rho}
\end{equation}

\noindent where $\rho$ is the ratio of the intensities carried by
the two interfering beams, $\rho =a_u^2/a_l^2$. The visibility $
{\mathcal{V}}$, which is a symmetric function of $a_u$ and $a_l$,
has the same value if $\rho$ is replaced by its inverse. A small
amplitude mismatch reduces the visibility but only very slightly,
as shown in figure \ref{theorycontrast}.

\subsection{Our main choices}

Our goal was to build an atom interferometer in which the two
atomic paths are sufficiently separated, so that one can apply a
perturbation to only one path: such an arrangement is necessary to
perform interferometric measurements of a perturbation. Previous
experiments of this type include the measurement of the electric
polarizability of sodium \cite{ekstrom95,roberts04} and the
measurement of the index of refraction of gases for sodium atomic
waves \cite{schmiedmayer95,roberts02}, both experiments being done
by D. Pritchard and co-workers. More recently, J. P. Toennies and
coworkers have compared the electric polarizability of helium and
helium dimer \cite{toennies03}. Moreover, we wanted to observe the
dependence of the index of refraction with the velocity of the
atomic wave and this dependence is detectable only if this
velocity is comparable to or larger than the thermal velocity of
the target gas. We have therefore chosen to use for the atomic
wave a thermal beam rather than a slow beam: this second choice
would have imposed to use also a cold atomic target, making the
experiment very complex.

We had to choose the diffraction process, among several
possibilities: diffraction by material gratings (which was first
studied by D. Pritchard and co-workers \cite{keith88} and briefly
reviewed in reference \cite{schollkopf04}), elastic diffraction by
a laser standing wave (first observed by Arimondo et al.
\cite{arimondo79}, with well resolved diffraction peaks first
recorded by D. Pritchard and co-workers \cite{moskowitz83}) or
inelastic diffraction processes, which can be either a one-photon
diffraction  process (used in Ramsey-Bord\'e interferometers)
\cite{riehle91}, or a two-photon Raman diffraction process used in
many cold or thermal atom interferometers (its first use being
described in reference \cite{kasevich91}). We have chosen to use
elastic Bragg diffraction by laser standing waves, the main
advantages being the high transmission of the interferometer
associated with a high fringe visibility and the fact that we can
use an ordinary single frequency laser. The first interferometers
using this diffraction process and thermal atoms were built by Siu
Au Lee and co-workers with metastable neon \cite{giltner95b} and
also by A. Zeilinger and coworkers using metastable argon (but not
in the Bragg regime) \cite{rasel95}. Elastic diffraction is
similar to diffraction by a material grating, in the sense that
the internal atomic state is not modified. The grating period is
equal to half the laser wavelength, which must be chosen very
close to a resonance transition of the atom, so that diffraction
can be observed with modest laser power densities. With sufficient
laser power densities, diffraction orders higher than the first
one can be easily observed \cite{martin88,giltner95a,koolen02}.

The choice of laser diffraction limits the choice of the atom
among those which have an intense resonance transition accessible
to cw single frequency lasers. If one excepts the use of
metastable states (with rare gases or hydrogen), this requirement
favors considerably the alkali atoms. Then, the most important
quantity is the Bragg angle $\theta_B= \lambda_{dB}/\lambda_L$,
which must be as large as possible in order to maximize the
separation between the two atomic paths in the interferometer.
Because the atomic de Broglie wavelength scales like $m^{-1}$, a
light atom is favored and we have chosen the lithium atom. Its
first resonance transition is at a $671$ nm wavelength,
corresponding to a grating period $a=335$ nm. By seeding lithium
in a supersonic beam of argon, the mean velocity $u$ of the
lithium atoms is close to $1060$ m/s corresponding to a de Broglie
wavelength $\lambda_{dB} = 54 $ pm and a Bragg angle $\theta_B=
80$ $\mu$rad.

\subsection{Elastic diffraction of atoms by a laser
standing wave}

As pointed out by Siu Au Lee and coworkers \cite{giltner95b},
diffraction in the Bragg regime is ideal to build an
interferometer: only two diffraction orders ($0$ and $p$) are
produced for a well chosen incidence angle and, by varying the
laser power density and/or the interaction time, the diffraction
efficiency can be tuned to produce $50-50$\% beam-splitters and
$100$\% mirrors. Therefore, the transmission of an ideal
Mach-Zehnder interferometer using this diffraction process should
be equal to $100$\% and, as a result of the symmetry of this
interferometer, the fringe visibility, measured on the $B_1$
output beam,  should also be equal to $100$ \%.

We first recall that elastic diffraction by a laser standing wave
results from the absorption of a photon going in one direction
followed by the stimulated emission of a photon going in the other
direction: this scheme corresponds to first order diffraction and
$p$ steps are needed for the diffraction order $p$. After an
absorption-emission cycle, the atom is back in its initial level,
and it has received a momentum kick equal to two photon momenta.

The laser frequency $\omega_L$ and the resonance atomic frequency
$\omega_0$ differ by the detuning $\delta = \omega_L -\omega_0$,
which must be considerably larger than the natural width $\gamma$
of the resonance transition. Then, the main effect of the laser
standing wave is to create a weak periodic potential proportional
to the local density of energy in the laser beam, $ V= V_0(z)
\cos^2(k_Lx)$ (where the $x$ axis is parallel to the laser beam
wave vector). The periodic nature of the potential can be treated
by introducing Bloch states as done in our previous paper
\cite{champenois01}, which quotes many previous works on laser
diffraction.

To simplify the discussion, we assume that $V(z)$ extends over a
distance $w$ (we do not define precisely $w$ which should not
confused with the Gaussian beam radius $w_0$ discussed below), so
that an atom with a velocity $v$ interacts with the laser beam
during an interaction time $t_{int} =w/v$. The natural energy unit
of the problem is the atomic recoil energy $\hbar\omega_{rec}=
\left(\hbar k_L\right)^2/(2m)$. Following
\cite{keller99,champenois01}, this quantity can be used to define
a dimensionless potential $q= V_0/(4 \hbar\omega_{rec})$ and a
dimensionless interaction time $\tau = \omega_{rec} t_{int}$.

The incident atom is characterized by its momentum state in the
$x$ direction, $\left| k_x \right>$. When $q$ is large ($q\gg 1$),
the periodic potential couples the incident atomic wave $\left|
k_x \right>$ to many other states $\left| k_x +2nk_L \right>$,
where $n$ is an integer. The Bragg regime occurs when $k_x \approx
\pm p k_L$ and if the potential $q$ is not too strong and does not
vary too rapidly with $z$. Then, one can neglect the coupling of
the two states $\left| k_x = \pm p k_L \right>$ with other states
and treat the dynamics as a two-level problem. At the lowest
nonvanishing order, the coupling between these two levels is
proportional to $q^p$ and the probability of diffraction of order
$p$ is then given by a Rabi oscillation:
\begin{equation}
\label{diffraction1}
 P_p = \sin^2\left( q^p \tau/d_p   \right)
\end{equation}
\noindent where the coefficient $d_p$ is equal to $1$ for order
$p=1$, to $4$ for order $p=2$ and to $64$ for order $p=3$. The
intensity which is not diffracted remains in the zeroth-order
beam. Because of the dependence in $q^p$ of the sine argument in
equation (\ref{diffraction1}), the $q$ values for a $50-50$\%
beam-splitter and a $100$\% reflective mirror are linked by
$q_{BS} = q_M \times 2^{-1/p}$. Finally, this diffraction process
induces some phase-shifts of the waves which will not discussed
here  but which may be very important \cite{buchner03}.

If $\delta$ is too small, real excitation of the atom followed by
a spontaneous emission of a photon occurs during the time spent by
the atom in the laser field. When this occurs, the coherence of
the atomic propagation is destroyed very efficiently. The
probability $P_{SE}$ of a spontaneous emission event is given by :

\begin{equation}
\label{diffraction2}
 P_{SE} = q \tau \frac{\gamma}{\delta}
\end{equation}

\noindent As $q\propto \delta^{-1}$ and $P_{SE} \propto
\delta^{-2}$, laser diffraction can be made almost perfectly
coherent by choosing a sufficiently large detuning. For a given
value of $q$, the use of a larger detuning requires also a larger
laser power density, so that the available laser power gives a
practical limit to the detuning.

\section{Experimental setup}

Our experimental setup is inspired by the sodium interferometer of
D. Pritchard and co-workers \cite{schmiedmayer97} and by the
metastable neon interferometer of Siu Au Lee and co-workers
\cite{giltner95b}. We are going to describe its main parts and to
explain our procedures to align its components.

\subsection{ Vacuum system}
The vacuum system is made of five differentially pumped chambers,
(see figure \ref{MZschematic}):

\begin{itemize}
\item the first chamber contains the supersonic beam source and is
pumped by a $8000$ l/s unbaffled oil diffusion pump (Varian
VHS400). The gas load due to the beam is a few mbar.l/s and, under
normal beam operation, the residual pressure is about $
8\times10^{-4}$ mbar. The beam exits this chamber through a $0.97$
mm diameter skimmer provided by Beam Dynamics.

\item the second chamber, which serves to differential pumping, to
collimation and to optical pumping of the lithium beam, is pumped
by a $2400$ l/s oil diffusion pump (Varian VHS6) with a water
cooled baffle. Under normal beam operation, the pressure is about
$3\times 10^{-6}$ mbar. The beam exits this chamber through the
source slit $S_0$.

\item the third chamber, which serves to collimation only, is
pumped by a $700$ l/s oil diffusion pump from Edwards with an
internal baffle. The residual pressure is below $5\times10^{-7}$
mbar, practically independent of beam operation. The beam exits
this chamber through the collimation slit $S_1$.

\item the fourth chamber, which contains the interferometer, is
pumped by two $1200$ l/s oil diffusion pumps (Varian VHS1200) with
water cooled baffles. The residual pressure is below
$5\times10^{-7}$ mbar. The detector slit $S_D$ is also located in
this chamber. The beam exits this chamber through a $3$ mm
diameter hole, located just before an UHV gate valve.

\item the fifth chamber holds the surface ionization hot-wire
detector. As the stray signal of such a detector is very sensitive
to the residual gas, this chamber is built with UHV components and
is pumped by a $300$ l/s turbo molecular pump. The residual
pressure in this chamber is a few $10^{-9}$ mbar, when the UHV
gate valve is closed and about $10^{-8}$ mbar when it is opened.
\end{itemize}

All the water baffles are cooled by circulating a liquid near
$3^{\circ}$C. We use three double stage roughing pumps: two 65
m$^3$/h pumps, one for the beam source, one for the other four oil
diffusion pumps and a $15$ m$^3$/h pump for the turbo pump of the
detector. To reduce vibrations induced in the setup, these pumps
are located in the next room.

\subsection{The atom wave source and detector}

Our lithium atomic beam, inspired by the design of Broyer, Dugourd
and co-workers \cite{blanc92} is briefly described in
\cite{delhuille02a,miffre04a,miffre04b} and more details will
appear in another paper. Lithium is seeded in argon and our normal
operating conditions are a source pressure of $330$ mbar, a
temperature equal to $973$ K for the back part of the oven (fixing
the lithium vapor pressure near $0.55$ millibar), a temperature
equal to $1073$ K for its front part and a nozzle diameter equal
to $200$ $\mu$m. We have measured the beam mean velocity, $u= 1060
$ m/s and the terminal parallel temperature of lithium
$T_{\|}\approx 6.6$ K. This parallel temperature is roughly $1/3$
of the calculated parallel temperature of the argon carrier gas,
an effect which occurs when a light species is seeded in a heavier
carrier gas \cite{miffre04a,miffre04b}.

To detect the output beams, we use a hot-wire detector which has
been fully described in a previous study \cite{delhuille02b}. Its
detection efficiency, which varies with the oxidation and the
temperature of the rhenium wire, was measured to be close to
$30$\%. With our normal operating conditions, the collimated beam
gives a signal up to $8 \times 10^4$ counts/second, on a
background signal close to $2\times 10^3$ counts/second. This
background signal presents a non-Poissonian statistics with a few
bursts.

On figure \ref{MZschematic}, it is clear that the location of the
detector must be well chosen. We must put the detector far enough
from the third laser standing wave, at a place where the two exit
beams $B_1$ and $B_2$ are well separated: these beams carry
complementary signals and the fringe visibility would be very
small if the detector was put close to the third grating, where
these two beams are strongly overlapping. The complementary
character of the two signals is a consequence of the fact that
laser diffraction is acting on the phase and not on the amplitude
of the atomic wave (for more details, see figure 7 of reference
\cite{champenois99}). However, we must not forget the existence of
the stray beams represented on figure \ref{MZschematic}. These
beams carry some flux, because the diffraction amplitudes are not
at their optimum values, and these stray beams cross the main exit
beams $B_1$ and $B_2$ at a distance equal to the inter-grating
distance $L_{12}= L_{23} = 0.605$ m. Therefore, we have chosen to
put the detector slit (which defines if an atom is detected or
not) at a distance $L_{34} = 0.40$ m from the third laser standing
wave, $0.2$ m in front of the place where these stray beams are
expected to create the largest signals. The hot wire itself is
$0.35$ m away in the fifth UHV vacuum chamber. In a first
arrangement, the detection slit, which was placed very near the
hot wire detector, was put out of order by excessive heating due
to the hot wire radiation.

\subsection{Laser standing waves}

We use an home-made single frequency cw dye laser, following F.
Biraben's design \cite{biraben82}, pumped by a Spectra-Physics
argon ion laser at $515$ nm. The dye is LD 688 from Exciton
dissolved in EPH. Using the H\"ansch-Couillaud \cite{hansch80}
frequency stabilization, we get a laser linewidth of the order of
$1$ MHz. The laser beam goes through a $60$ dB optical isolator.
After the isolator, the power available at $671$ nm is close to
$400$ mW, for $5$ W of Ar$^+$ pump power.

The laser frequency, which is measured by a home-made lambdameter,
must be detuned from resonance, which has a complex structure due
to the fine, hyperfine and isotopic splittings
\cite{sansonetti95}. Most of our experiments are optimized for the
$^7$Li isotope (natural abundance $92.5$\%) and we define the
frequency detuning by:

\begin{equation}
\label{diffraction2} \delta/(2\pi) = \nu_L - \left(E(^2P_{3/2}) -
E(^2S_{1/2}, F= 1)\right)/h
\end{equation}
\noindent where the energies are those of the $^7$Li isotope
levels. The hyperfine structure of the $^2P_{3/2}$ state is very
small and can be neglected. Our usual choice of detuning is
$\delta/(2\pi) =+ 3.0$ GHz and whenever a different value is used,
it will be indicated. The natural width of the $^2S_{1/2}$ -
$^2P_{3/2}$ transition of lithium is $\gamma/2\pi =$ 5.9 MHz
\cite{mcalexander96}.

The laser beam is magnified by a telescope made of two AR coated
singlet lenses so that we can change the magnifying ratio by
changing one lens. We characterize the beam transverse profile by
scanning a photodiode through it, thanks to a motorized
translation stage, and the recorded intensity as a function of the
photodiode position is fitted to a Gaussian profile, thus
extracting the Gaussian beam radius $w_0$. When operating with low
power densities (practically only when using first order
diffraction), the Gaussian beam is limited by an iris and the
resulting beam profile is closer to a flat top profile.

The beam is then split by two beam splitters with a nominal
transmission equal to $50$\% for an incidence of $45^{\circ}$. We
thus get three beams, one with a power close to $P/2$ and two
beams, each with a power close to $P/4$. The $P/2$ beam serves for
the central laser standing wave, on mirror $M_2$, while the two
$P/4$ beams serve for the other laser standing waves, on mirror
$M_1$ and $M_3$. Using incidence angles different from
$45^{\circ}$, we are able to modify the power repartition between
these three beams: this is needed when using first order
diffraction because the real transmission differs from $50$\% and
also when using higher diffraction orders $p=2$ and $3$, because
the needed power repartition is not the same. In order to choose
the best laser power repartition, we have recently installed
attenuators made of an half-wave plate followed by a polarizer on
two of these three laser beams. This system was not available
during most of the experiments described here.

The three laser beams are sent near normal incidence on the
mirrors $M_j$. The properties of a standing wave are weakly
sensitive to the exact value of the incidence angle on the mirror
and very sensitive to the orientation of the direction
perpendicular to the mirror surface. More precisely, if a plane
wave is incident on a mirror with a small angle of incidence $i$,
the reflected wave and the incident wave produce a wave which is
progressive in a direction parallel to the mirror surface, with a
wave vector $k_L \sin i$ and which is a standing wave in the
direction normal to the mirror with a wave vector $k_L \cos i$.
The progressive character of the wave parallel to the mirror
surface induces a Doppler shift of its frequency equal to $k_Lu
\sin i$ which corrects the detuning: in our experiment, this
Doppler shift of the order of $1.5$ MHz per mrad is perfectly
negligible. The fact that the laser wave vector normal to the
mirror surface is $k_L \cos i$ slightly modifies the momentum kick
received by the atoms which becomes $2 p k_L \cos i$ but, for
small $i$ values, this modification is negligibly small.

Following equation (\ref{g7}), the phase of the interference
fringes depends on the $x$-positions of the three mirrors and this
property makes the interferometer very sensitive to vibrations. In
the interferometers developed by D. Pritchard
\cite{schmiedmayer97} and by Siu Au Lee \cite{giltner95b}, these
vibrations were controlled by servo-loops. We have chosen to
minimize these vibrations by building a very rigid rail to support
the three mirrors $M_j$. This rail and the role of vibrations will
be discussed in an other paper. As in references
\cite{schmiedmayer97,giltner95b}, we use an optical three-grating
Mach-Zehnder interferometer to control the vibrations of the
$x$-positions of the three mirrors $M_j$ and the measured noise on
the quantity $\left(2 x_2 -x_1 -x_3 \right)$ is negligibly small,
with a rms amplitude of the order of a few nanometers. The output
signal of this interferometer is also used to calibrate the
displacement of the motion of the mirror $M_3$, which serves to
observe interference fringes.

\subsection{Alignment procedures}

We must align the atomic beam and the mirrors producing the laser
standing waves. The numerous adjustments must be done with great
care: to give an idea, most angles must finally be tuned within
about $10$ $\mu$rad from their optimum value.

The atomic beam alignment is difficult as the beam must go with
minimum attenuation from the nozzle to the hot-wire of the
detector $3.4$ m away, through the skimmer, the source slit $S_0$,
the collimation slit $S_1$, the detector slit $S_D$, the $3$ mm
diameter hole located before the detector chamber. For each
element, we explain the available adjustments and how we proceed
to make them:

\begin{itemize}

\item the oven can be adjusted in the three directions under
operation.

\item the skimmer and the $3$ mm diameter hole are fixed to the
center of their supporting flanges, while all the other elements
can be adjusted in the ${\mathbf x}$ direction, but not in the
${\mathbf y}$ direction. This is possible because the three slits
have a sufficient height, about $10$ mm.

\item the width of the slit $S_0$ is fixed and equal to $20$
$\mu$m, while the widths of the collimation slit $S_1$ and of the
detector slit $S_D$ are controlled by piezo-drives from
Piezosystem Jena in the $0-200$ $\mu$m range: the slit widths
commonly used are $e_1 = 12$ $\mu$m for $S_1$ and $e_D = 50$
$\mu$m for $S_D$ (if different values are used, they will be
specified). The slit material has been chosen to be non magnetic,
because the inhomogeneous field which would exist in the slit
opening could induce a spreading of the atomic beam, by Stern and
Gerlach effect.

\item the slits $S_0$, $S_1$ and $S_D$ are made vertical before
operation. We have used either the diffraction pattern of a laser
beam or the observation of the slit with a telescope, in
comparison with a plumb line. We estimate that the slits are
vertical within a few mrad: if the useful height of the slit $S_i$
is $h_i$, a small error $\epsilon$ on its verticality induces no
broadening of the full width at half maximum of the beam but a
broadening of its wings of the order of $\epsilon h_i$. We can
evaluate these useful heights simply by assuming straight lines
trajectories for the atoms, from the skimmer to the $3$ mm hole
near the detector. The calculated useful height is $1.3$ mm range
for $S_0$, $1.8$ mm for $S_1$ and $2.9$ mm for $S_D$. The
corresponding broadening of the beam wings, of the order of $1-3$
$\mu$m per mrad, is probably fully negligible for $S_0$ and $S_D$
which are rather wide, and less negligible for $S_1$ which has
usually the smallest width.

\item the $x$-position of these three slits and of the hot-wire
can be modified under vacuum: in each case, we use a translation
stage operated by a linear drive vacuum feedthrough, with a
sensitivity of the order of $10$ $\mu$m. In addition, the
$x$-position of the detector slit $S_D$ can be swept under
computer control by a piezo-translation from Piezosystem Jena over
$400$ $\mu$m.

\end{itemize}
For the laser standing waves, each mirror $M_j$ is attached to a
double stage kinematic mount built in our laboratory. The first
stage, with screws, can be operated only when the experiment is at
atmospheric pressure while the second stage actuated by
low-voltage piezo-translators, has a tuning range close to $600$
$\mu$rad. A first alignment, within $\pm 100$ $\mu$rad, must be
made with the experiment at atmospheric pressure and the final
tuning is made with the second stage. To make the first alignment,
we use the following signals:
\begin{itemize}

\item  we first adjust the rotation $\theta_z$ around the
horizontal axis ${\mathbf z}$ with an autocollimator. For each
mirror $M_j$, we set the autocollimator by observing the
horizontal surface of diffusion pump oil through a pentaprism and
we set the mirror perpendicular to the autocollimator axis.

\item we then adjust the rotation $\theta_y$ around the vertical
axis ${\mathbf y}$, with a laser beam which replaces the atomic
beam, going from the skimmer to the $3$ mm hole near the detector.
Then, using a pentaprism, we send this beam successively on each
mirror $M_j$ and we set the mirror so as to maximize the reflected
laser power measured behind the skimmer.
\end{itemize}

With the experiment under vacuum, we make the final adjustment of
$\theta_y$ for each mirror $M_j$: we tilt the mirror to observe
Bragg diffraction of the chosen order $p$ (see figure
\ref{diffraction}) with the corresponding laser standing wave. We
have no signal which can be used to finely tune the $\theta_z$
angles separately, but we must tune one of these three angles to
cancel $\Delta {\mathbf k}_G$ defined by equation (\ref{g61}) and
the fringe visibility is very sensitive to an exact cancellation,
as shown below in figure \ref{deltak}. We use mirror $M_2$ as its
effect is twice as large as the effect of $M_1$ or $M_3$.

Finally, an optical grating is linked to each mirror $M_j$ to form
the optical three-grating Mach-Zehnder interferometer briefly
discussed above. It is necessary to align this interferometer
before the final adjustments of the mirrors $M_j$.

\section{Atom interference effects}

\subsection{Diffraction experiments}

\begin{figure}[hbt]
  \begin{center}
  \includegraphics[width= 11.5 cm]{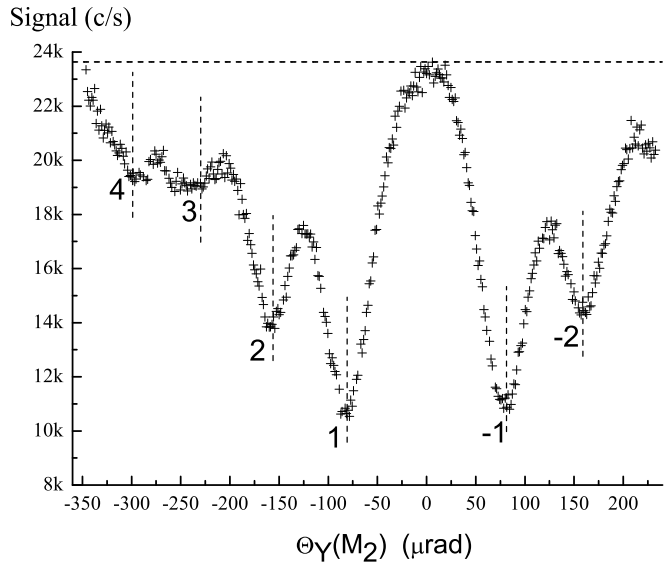}
\caption{\label{diffraction} Intensity of the direct beam measured
as a function of the angle $\theta_y$ of mirror M2. When the Bragg
condition is fulfilled for a diffraction order $p$, the
transmitted intensity goes through a minimum labelled by the order
$p$. This experiment was made with an almost Gaussian laser beam
with a measured waist radius $w_0 = 3.1$ mm, a power $P = 240$
milliwatts and a detuning $\delta/(2\pi) = 1.2$ GHz. The
collimation slit width was $e_1 = 10$ $\mu$m and the detection
slit was $e_D = 70$ $\mu$m.}
  \end{center}
  \end{figure}

With only one laser standing wave, we can observe diffraction. Two
main types of diffraction experiments have been done:
\begin{itemize}
\item by setting the orientation of the mirror in order to be in
the Bragg geometry, we produce a diffracted beam. Then, by
scanning the position of the detector slit, we can record the
profile of the direct and diffracted beam. A typical result was
shown in our previous paper \cite{delhuille02a}. We have verified
that the diffraction behaves as expected in the Bragg regime,
with, in particular, the absence of a beam of order $p=-1$ when
the geometry favors the diffraction of order $p=1$.

\item by rotating the mirror around the $y$-axis, we successively
fulfill the Bragg condition for the various diffraction orders
$p$. We have recorded the direct beam intensity as a function of
the angle $\theta_y$ and diffraction then appears as an intensity
loss. Figure \ref{diffraction} presents such a recording. We
observe intensity losses corresponding to Bragg condition for the
orders $p=-2$ up to $p=4$. The interest of such a recording is
that it gives immediately an idea of the diffraction efficiency.
We never reach a $100$\% diffraction probability, because of the
presence of $^6$Li and of the finite widths of the velocity and
angular distributions of the incident atomic beam.
\end{itemize}

With our usual detuning $\delta/(2\pi) = 3.0$ GHz and with the
typical power density used for first order diffraction, the
diffraction probability for the $^6$Li atoms is very small and we
may forget their presence. We have also made some experiments with
a detuning chosen to diffract selectively these atoms.

\subsection{High visibility atom interference fringes}

\begin{figure}[p]
  \begin{center}
  \includegraphics[width= 9.5 cm]{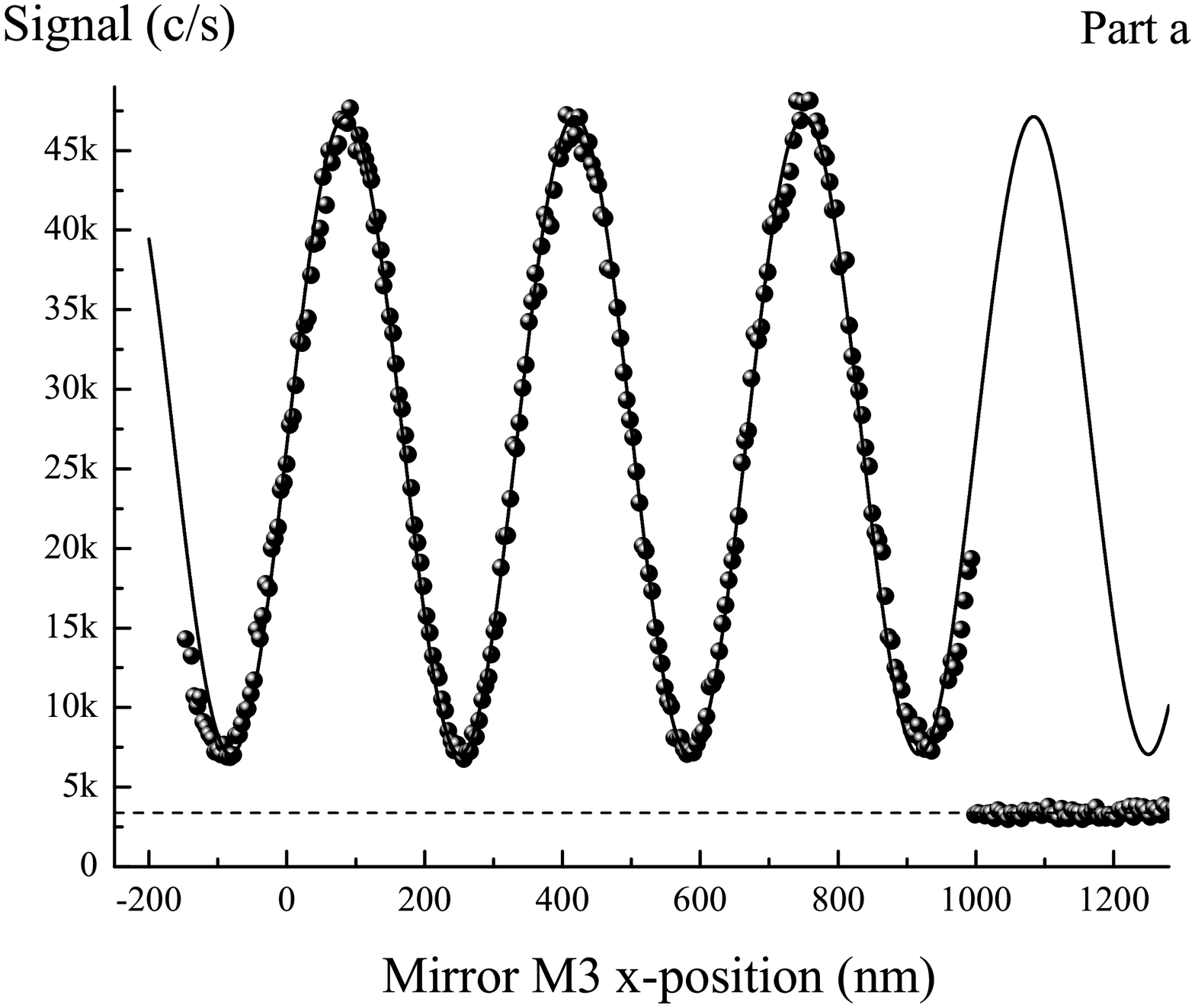}
  \includegraphics[width= 9.5cm]{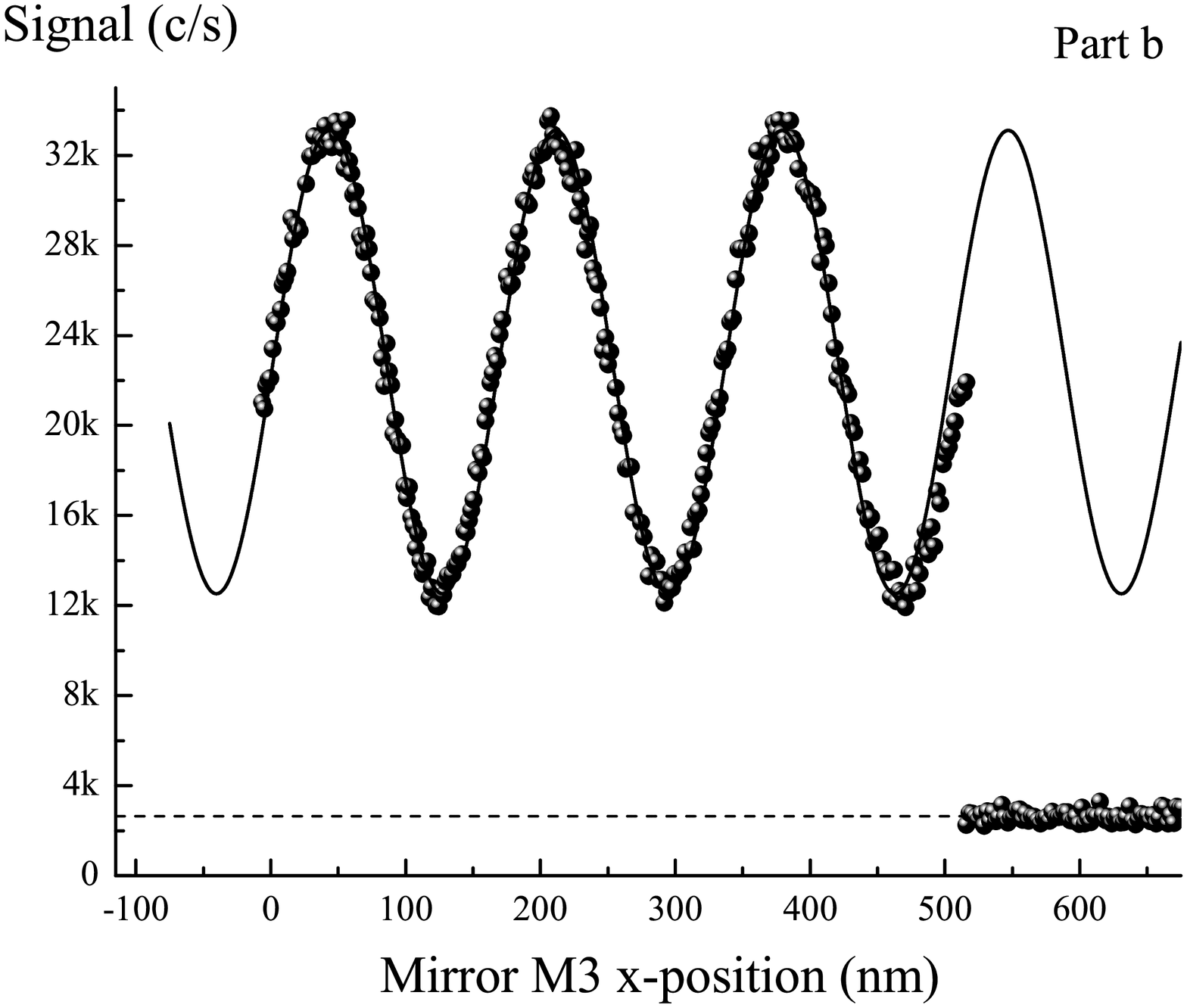}
  \includegraphics[width= 9.5 cm]{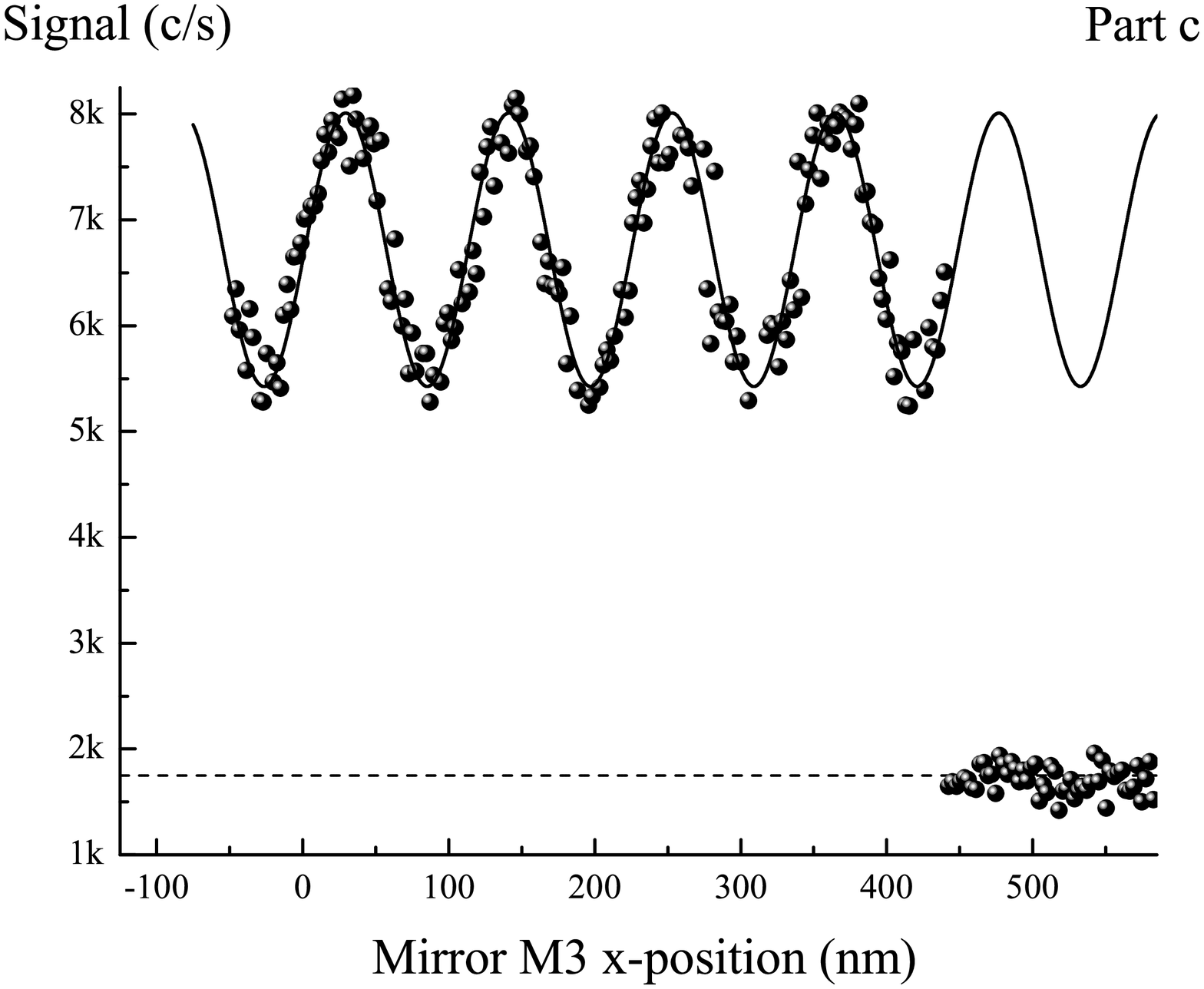}
\caption{\label{atomicfringes}Interference signals recorded with
the diffraction orders $p=1$ (part a, $84.5$ \% visibility,
collimation slit width $e_1 = 12$ $\mu$m, detection slit width
$e_D = 40$ $\mu$m), $p=2$ (part b, $51$\% visibility, $e_1 = 14$
$\mu$m, $e_D = 50$ $\mu$m) and $p=3$ (part c, $26$ \% visibility,
$e_1 = 12$ $\mu$m, $e_D = 40$ $\mu$m). In these three cases, the
$B_1$ output signal is measured as a function of the $x$-position
of mirror $M_3$, calibrated thanks to the optical interferometer
linked to the three mirrors. The counting time is equal to $0.1$ s
and one can see that the displacement $\Delta x$ necessary to
sweep one fringe is equal to $\lambda_L/(2p)$. The background
signal recorded just after the recording of the signal is also
plotted.}
  \end{center}
  \end{figure}

We have operated our interferometer using successively three
different diffraction orders $p=1$, $2$ and $3$. By sweeping the
$x$-position of mirror $M_3$, we have observed interference
signals with a very high visibility which are plotted in figure
\ref{atomicfringes}. In all cases, the signal is expressed as a
number of atoms detected per second with an usual counting time
equal to $0.1$ s. The observed signal can be written as:

\begin{equation} \label{e0}
I_1 = I_B + I_0 \left[1+ {\mathcal{V}}\cos\phi \right]
\end{equation}
\noindent The background signal $I_B$ of the detector is recorded
just after or before recording the signals, by flagging the atomic
beam in the second chamber and we deduce from this measurement the
mean $I_B$ value. Then, we can make a fit of the signal to
estimate the mean intensity $I_0$ and the visibility
${\mathcal{V}}$. The phase $\phi$ is a locally linear function of
time, but the fit must take into account the nonlinearity of the
piezo drive. Table ~\ref{expvalues} gives for the three orders $ p
= 1$, $2$ and $3$ the parameters used (laser detuning, beam waist
$w_0$ and beam powers) and the mean intensity $I_0$ and the
visibility ${\mathcal{V}}$ deduced from the fits.

 We have measured the interferometer transmission by making the
ratio of the intensity at the peak of constructive interference
and of the intensity of the direct atomic beam, in the absence of
the three laser standing waves. With first order diffraction, the
measured transmission can reach quite large values, up to $85$\%.
Theory predicts a $100$\% value and the difference is due mainly
to imperfections of the diffraction process and to the presence of
$^6$Li in the beam with its natural abundance equal to $7.5$\%.

\begin{table}[htb]
\caption{\label{expvalues} This table collects information
concerning our best signals obtained with the diffraction order
$p$. We give the date of the experiment, the mean intensity $I_0$,
the visibility $\mathcal{V}$ and several experimental parameters:
the Gaussian beam radius $w_0$, the laser detuning
$\delta/(2\pi)$, the total laser power $P$ used in the laser
standing waves, the collimation slit width $e_1$ and the detection
slit width $e_D$. We have also calculated the figure of merit
$I_0\mathcal{V}^2$, related to the phase sensitivity if Poissonian
statistics is assumed.}

 \begin{center}
 \begin{tabular}{|c|c|c|c|c|c|c|c|c|c|}
  \hline
   $p$  & Date & $I_0$ (c/s) & $\mathcal{V}$ \% & $I_0\mathcal{V}^2$  & $w_0$ (mm) & $\delta/(2\pi)$ (GHz) & $P$(mW) & $e_1 (\mu$m) & $e_D (\mu$m) \\
  \hline
  \hspace{0.15 cm}     1 &   March 2004     &  12900 & 80.5 $\pm$ 1 & 8360  & 5.0 (a) & 2.8 & 150 & 12 & 40 \\
  \hspace{0.15 cm}       &   July 2004  (b) &  23710 & 84.5 $\pm$ 1 & 16930 & 5.0 (a) & 2.8 & 150 & 12 & 40 \\
  \hline
  \hspace{0.15 cm}     2 &   April 2004     &  14430 & 49.0 $\pm$ 1 & 3465  & 2.9     & 1.5 & 300 & 12 & 50 \\
  \hspace{0.15 cm}       &   Sept. 2004     &  20180 & 51.0 $\pm$ 1 & 5250  & 1.8     & 3.1 & 460 & 14 & 50 \\
  \hspace{0.15 cm}       &   Sept. 2004 (b) &   8150 & 54.0 $\pm$ 1 & 2735  & 1.8     & 3.1 & (c) & 14 & 60 \\
  \hline
  \hspace{0.15 cm}     3 &   April 2004     &  4870  & 26.0 $\pm$ 1 &  304  & 2.9     & 1.1 & 300 & 12 & 40 \\
  \hline
  \end{tabular}
 \end{center}

{(a) when using $p=1$, the intensity profile has a flat top and
$w_0$ is the radius of the laser beam}

{(b) experiment done with a cancellation of the effect of the
magnetic field gradient}

{(c) not measured during the experiment}

  \end{table}

The dependence of the fringe visibility with the diffraction order
has been studied only once before, by Siu Au Lee and coworkers
\cite{giltner95b}: in this experiment like in the present case,
the visibility decreased rapidly with increasing order:
${\mathcal{V}} =62$\% for $p=1$, ${\mathcal{V}} =22$\% for $p=2$
and ${\mathcal{V}} =7$\% for $p=3$. The most natural explanation
of this rapid decrease is the existence of a phase noise with an
amplitude proportional to the diffraction order $p$: this is the
case if the phase noise comes from the grating vibrations.
However, two other effects may also contribute to the rapid
decrease of the fringe visibility when the order $p$ increases:
\begin{itemize}

\item the incoherent processes involving a real photon absorption
followed by spontaneous emission are not negligible with the power
densities used for orders $p=2$ or $3$

\item the diffraction phase-shifts \cite{buchner03} behave like
$q^2 \tau$ and may be rather large during the diffractions of
orders $p=2$ or $3$. A large phase shift does not induce a loss of
fringe visibility if it is the same for all the atoms. The
dependence of the phase shift with time (due to the intensity
fluctuations of the laser), with space (due to the intensity
profile of the laser beams) and with the atom velocity may result
in a large reduction of the fringe visibility.

\end{itemize}

We think that decoherence by collision with the residual gas is
negligible in our case. This decoherence effect, which has been
studied in a Talbot-Lau interferometer with fullerenes
\cite{hornerger03,hornberger03a,hackermuller03}, could be observed
in our case if a lithium atom can be detected with a large
probability even after a collision with an atom of the residual
gas. Obviously, this is not the case. The residual gas creates an
index of refraction proportional to its density and the
transmitted waves are attenuated and phase shifted
\cite{schmiedmayer95,roberts02}. The fluctuations of these phase
shifts could induce a phase noise and a reduction of the fringe
visibility, but this effect is negligible in our experiment.
Moreover, this decoherence effect has no strong dependence with
the diffraction order $p$.

By moving the detector slit, we have successively recorded the
interference signals on the two outputs beams, $B_1$ and $B_2$,
and we have verified that a destructive interference at $B_1$
corresponds to a constructive interference at $B_2$. The observed
visibility at $B_2$ is slightly less good than at $B_1$: the
simplest explanation, which would be that the two interfering
beams have not equal amplitudes, is not convincing (see equation
(\ref{g8}) and figure \ref{theorycontrast}). We think that the
visibility difference is most probably due to the stray beams
represented in figure \ref{MZschematic}.

We have also been able to observe signals due to the $^6$Li
isotope present in the lithium beam with its natural abundance
($7.5$\%). This was done by changing the laser frequency so that
the diffraction was isotopically selective in favor of $^6$Li: for
this experiment, we used a laser frequency with a detuning of
$\delta/(2\pi) \approx -24$ GHz, so that the laser is at $4$ GHz
on the red side of the $^2S_{1/2}$ - $^2P_{1/2}$ transition of the
$^6$Li isotope and at $14$ GHz on the red side of the nearest
transition of the $^7$Li isotope, which is also the $^2S_{1/2}$ -
$^2P_{1/2}$ transition. We thus observe a mean intensity $I_0= 4
240$ s$^{-1}$ and a visibility ${\mathcal{V}} = 55$ \%.
Considering the $7.5$\% natural abundance of $^6$Li, the observed
mean intensity is too high to be purely $^6$Li; we think that a
noticeable contribution comes from the $^7$Li content of various
stray beams (with the detuning used, the probability of
diffraction of $^7$ Li atoms by one of the three laser standing
waves is small but not fully negligible). As these stray beams
carry no interference effect, their contribution to the signal
could explain a too large value of the mean intensity and, at the
same time, a visibility which is smaller than what we observe with
when we work with $^7$Li.

\section{Optimization of the fringe visibility}

We have explored how the defects modify the fringe visibility in a
systematic way. These effects can be analyzed theoretically
\cite{turchette92,champenois99} and we will compare the results of
this analysis with our experimental results.

\subsection{Sensitivity of the visibility to the orientations of the
standing wave mirrors}

We have not made any systematic study of the effect of the
rotations around the ${\mathbf y}$ axis: these rotations modify
the angle of incidence of the atomic wave on the laser standing
wave. When this angle differs sufficiently from the Bragg angle,
the diffraction  amplitude is reduced. The output signal and the
fringe visibility should also be reduced, but, following equation
(\ref{g8}), the associated visibility reduction is expected to be
very slow. On the contrary, the rotation around the ${\mathbf z}$
axis has a very large effect, as explained by the simple plane
wave theory recalled in paragraph II.A. The two waves, which
interfere on the detector, present a wave vector difference equal
to:

$$ \Delta{\mathbf k} = p\left(2{\mathbf k}_{G2} - {\mathbf
k}_{G1}-{\mathbf k}_{G3}\right)$$

\noindent The signal comes from the integration over the detector
surface of the local intensity.  If we assume that a flat
intensity profile over a region $-h_D/2 <y<h_D/2$ and zero
intensity elsewhere, we calculate a visibility given by:
\begin{equation} \label{e3}
{\mathcal{V}} = {\mathcal{V}}_0 \left| \mbox{ sinc} \left(\Delta
k_y h_D \right)\right|
\end{equation}
\noindent where $\mathcal{V}_0$ stands for the visibility achieved
when $\Delta {\mathbf k}=0$, $\Delta k_y$ is the $y$ component of
$\Delta{\mathbf k}$ and sinc$(x)$ is a short-hand notation for
$\sin (x)/x$.

\begin{figure}[hbtp]
  \begin{center}
   \includegraphics[width= 11.5 cm]{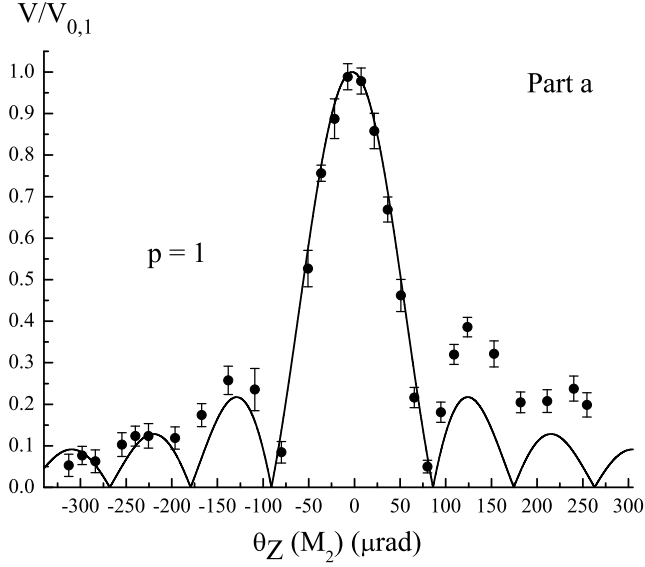}
  \includegraphics[width= 11.5 cm]{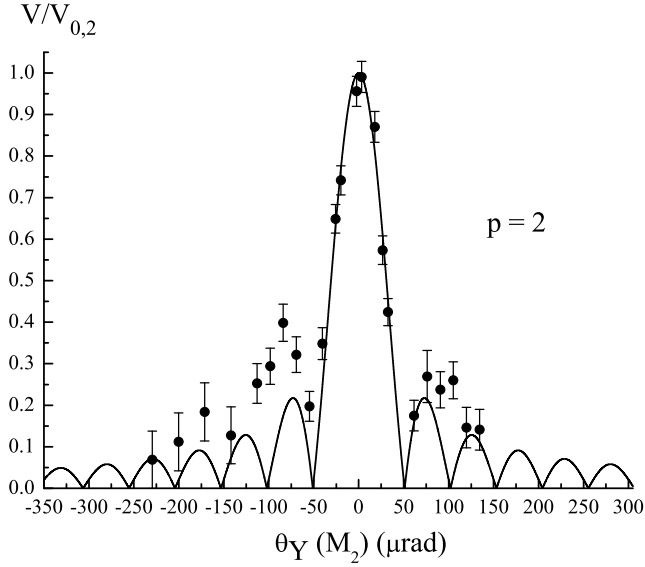}
\caption{\label{deltak} Fringe visibility measured as a function
of the angle $\theta_z$ measuring rotation around the ${\mathbf
z}$-axis of mirror $M_2$. The experiment has been done with the
diffraction orders $p=1$ (part a) and $p=2$ (part b). The points
are experimental and the curves are the best fits using equation
(\ref{e3}). The agreement is excellent in the central region,
where the visibility decreases twice as fast as when using $p=2$
than when using $p=1$, in agreement with equation (\ref{e3}).}
  \end{center}
  \end{figure}

We have tilted mirror $M_2$ around the ${\mathbf z}$-axis, by
applying a voltage on the corresponding piezo-drive and we have
recorded fringes and measured their visibility. We have converted
the voltage applied on the piezo-drive into a rotation angle,
using an external calibration and neglecting the piezo hysteresis.
The measured visibility has been plotted as a function of the
angle $\theta_z(M_2)$ in figure (\ref{deltak}). The visibility
decreases rapidly, as expected, but it does not vanish where
predicted by equation (\ref{e3}). We think that this is a kind of
apodization effect: the predicted cancellations disappear if a
smooth weight function of $y$ replaces the $0$ or $1$ intensity
function used to establish equation (\ref{e3}). D. Pritchard and
co-workers have made a study very similar to the present one in
\cite{schmiedmayer97}.

\subsection{Fringe visibility as a function of the mismatch between
the distances between consecutive gratings}

If the distances between consecutive gratings $L_{12}$ and
$L_{23}$ are different, the symmetry of the Mach-Zehnder
interferometer is broken and the visibility is reduced. This
effect was studied by numerical simulation by Turchette and
coworkers \cite{turchette92}. We have shown \cite{champenois99}
that, if the diffraction due to the slit $S_1$ is negligible, the
visibility is given by:

\begin{equation} \label{e4}
{\mathcal{V}} = {\mathcal{V}}_0 \left| \mbox{ sinc}
\left(\frac{pk_G e_0 \Delta L}{2L_{04}}\right) \mbox{ sinc}
\left(\frac{p k_G e_D \Delta L}{2L_{04}}\right)\right|
\end{equation}
\noindent where $\Delta L = L_{23} - L_{12}$ (this formula was
written in \cite{champenois99} for the diffraction order $p=1$
only).

\begin{figure}[hbtp]
  \begin{center}
  \includegraphics[width= 10 cm]{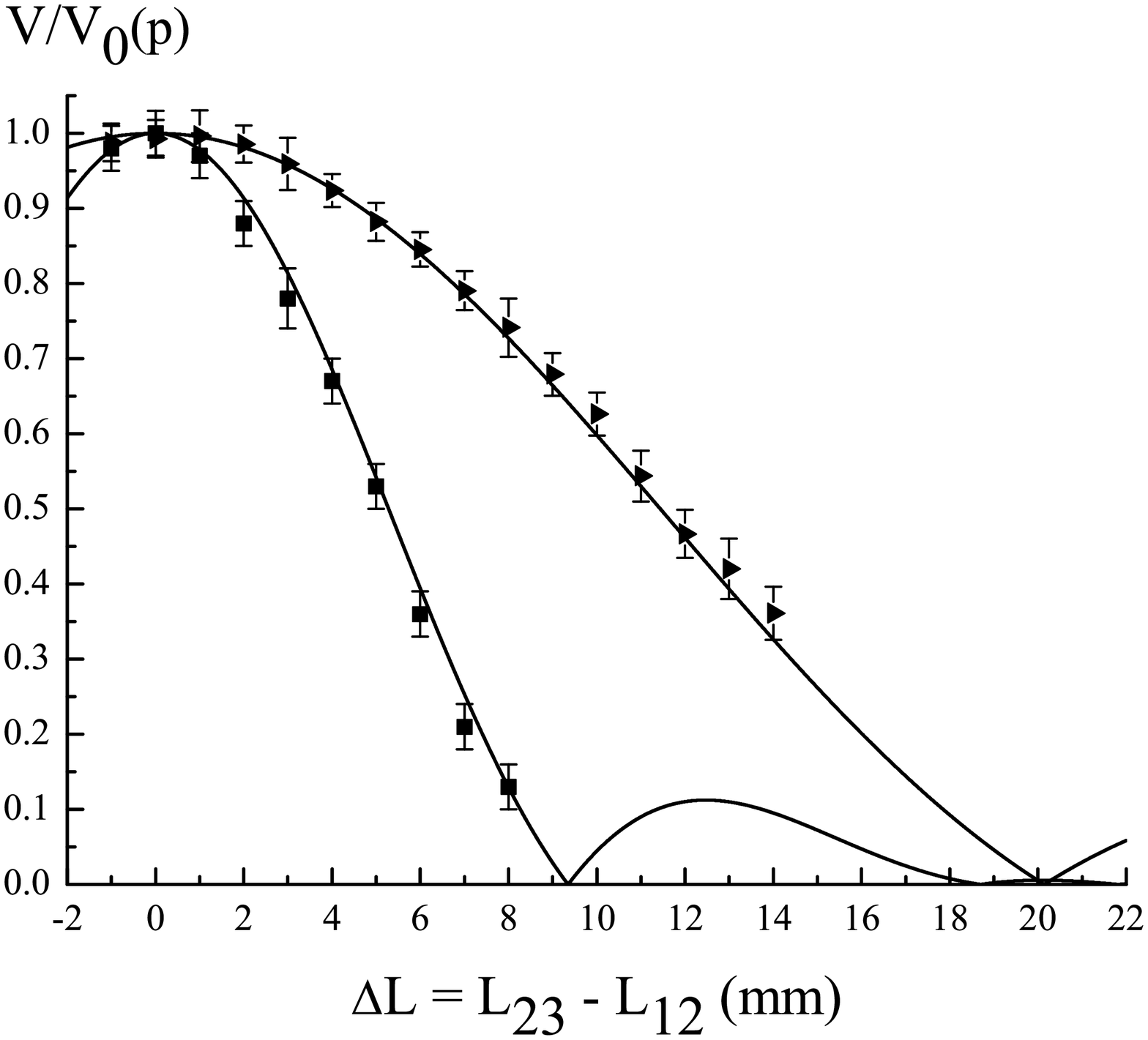}
\caption{\label{distancemismatch} Fringe visibility as a function
of the mismatch between the distances between consecutive gratings
$ \Delta L = L_{23}- L_{12} = z - z_c$ for first (triangles) and
second (squares) diffraction orders. The points are experimental
and the curve is the best fit using our approximate formula
(equation (\ref{e4})), for a collimation slit width $e_1 = 14$
$\mu$m. The fitted parameter are the maximum visibility ${\mathcal
V}_{0}(p)$ for each order $p$ and the position $z_c$ corresponding
to a vanishing distance mismatch.}
  \end{center}
  \end{figure}

To study this effect, we have moved the last mirror encountered by
the laser beam on its way to the mirror $M_1$ where it reflects
and forms the first laser standing wave. This motion was done with
a translation stage, so that the laser beam direction is
conserved. For various positions $z$ of this translation stage, we
have recorded atom interference signals and measured their fringe
visibility ${\mathcal{V}}$. The measurements have been fitted by
equation (\ref{e4}), in which we have replaced $\Delta L = z-z_c$,
where $z_c$ corresponds to the position which cancels the mismatch
$\Delta L$. The data points and their fit are plotted in figure
\ref{distancemismatch} and the agreement is very good. We cannot
explore a larger range of $z$ values because of the limited window
diameter. By a direct measurement on our machine, we have verified
that the value of $z_c = 3.5$ mm deduced from the fit corresponds
well, with an uncertainty of $\pm 0.5$ mm, to the equality of the
two distances $L_{12}$ and $L_{23}$.

\subsection{Signal and fringe visibility as a function of slit widths}

\begin{figure}[hbtp]
  \begin{center}
  \includegraphics[width= 11.5 cm]{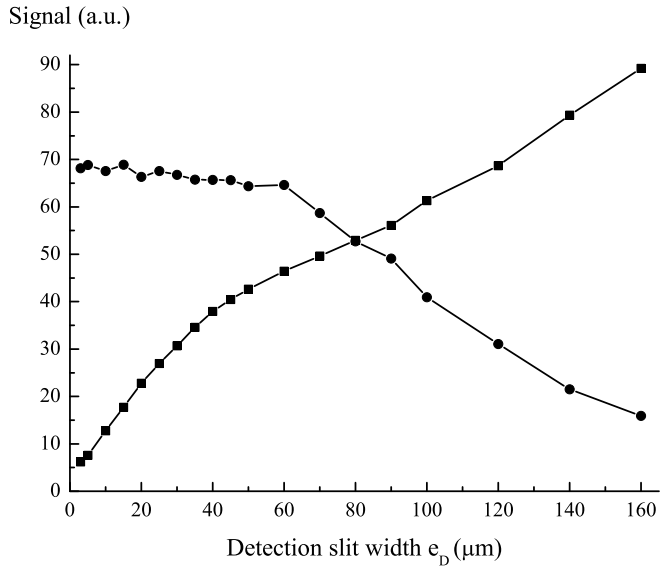}
\caption{\label{detectorslit} Fringe visibility ${\mathcal V}$ in
\% (dots), and mean signal intensity $I_0$ in $10^3$ counts/s
(squares) as a function of the detector slit width $e_D$ in $\mu$m
while the collimation slit width is $e_1 = 12$ $\mu$m. The lines
are simply drawn to guide the eye.}
  \end{center}
  \end{figure}

\begin{figure}[hbtp]
  \begin{center}
  \includegraphics[width= 11.5 cm]{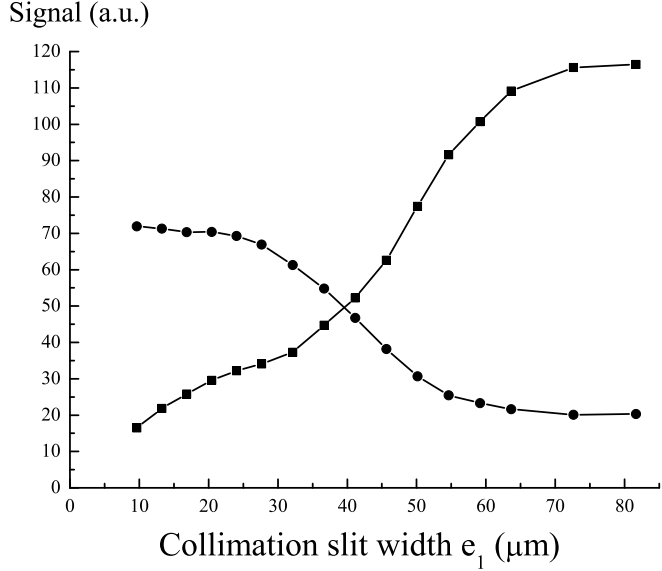}
\caption{\label{collimationslit} Fringe visibility ${\mathcal V}$
in \% (dots) and mean signal intensity $I_0$ in $10^3$ counts/s
(squares)as a function of the collimation slit width $e_1$ in
$\mu$m while the detection slit width is $e_D = 43$ $\mu$m. The
lines are simply drawn to guide the eye.}
  \end{center}
  \end{figure}

The widths of the collimation and detector slit can be adjusted by
piezo actuators and they open symmetrically. We have varied these
two slit widths and we have recorded the interference signals on
which we have measured the fringe visibility ${\mathcal{V}}$ and
the mean intensity $I_0$. These two quantities are plotted as a
function of the detector slit width $e_D$ in figure
\ref{detectorslit} and as a function of the collimation slit width
$e_1$ in figure \ref{collimationslit}. This study is very useful
to optimize the phase sensitivity of the interferometer.

If we consider first figure \ref{detectorslit} representing the
effects of the detector slit width $e_D$, the signal intensity
$I_0$ increases linearly with $e_D$ up to $e_D \approx 40$ $\mu$m
while the visibility ${\mathcal V}$ is roughly constant as long as
$e_D <60$ $\mu$m: this first regime is what is expected when the
detector slit collects only the signal corresponding to beam
$B_1$. Then for larger $e_D$ values, the intensity $I_0$ increases
more slowly and the visibility ${\mathcal V}$ decreases rapidly.
Now, the detector slit is sufficiently opened to collect all the
$B_1$ beam and a part of the $B_2$ beam. If the interferometer was
perfectly symmetrical, the $B_2$ beam would carry the same flux as
$B_1$ beam with a complementary interference signal. The fact that
the intensity increases with a slope reduced roughly by a factor
$2$ is in agreement with the fact that the slit opens
symmetrically and only one side of the slit is useful to transmit
the $B_2$ beam and the rapid decrease of the visibility is in good
agreement with the fact that the two beams $B_i$ carry
complementary interference signals.

When the collimation slit width $e_1$ is varied, the effects are
slightly more complex. In particular, one should not forget that
Bragg diffraction has a strong angular selectivity: this
selectivity makes that when the slit is widely opened, it admits
in the interferometer atoms which have not the Bragg incidence and
therefore these atoms have a low diffraction probability. These
atoms will contribute to make the direct stray beam (the beam
which is diffracted three times in the zeroth order) more intense.
As long as the collimation slit width $e_1$ is below $25$ $\mu$m,
the intensity $I_0$ increases with the slit width while the
contrast is mostly constant. When $35< e_1 <70 $ $\mu$m, the
intensity increases more rapidly, as a consequence of the
broadening of the wings of the direct beam. As the direct beam
carries no interference signal, the visibility decreases while the
product $I_0 {\mathcal V}$ remains roughly constant. Finally, when
$e_1> 70$ $\mu$m, the intensity saturates because the direct stray
beam fully covers the detection slit and the visibility remains
constant.

\subsection{Fringe visibility as a function of an applied magnetic field
gradient}

An atomic Mach-Zehnder interferometer operating with paramagnetic
atoms like lithium remains insensitive to a weak homogeneous
magnetic field but the output signal is very sensitive to a
magnetic field gradient, as explained below. This effect was
studied by D. Pritchard and co-workers
\cite{schmiedmayer94,schmiedmayer97} and also by D. Giltner in his
thesis \cite{giltner96}.

In our experiment, the Earth magnetic field is not compensated.
Moreover, the vacuum pipes are supported by a very heavy structure
made of steel rails, but we have made efforts to use very few
magnetic parts inside the interferometer vacuum chamber, the only
exception being small steel springs in the kinematic mounts of the
three mirrors. The field along the atomic paths has been measured,
it is very roughly homogeneous, reminiscent of the Earth field (of
the order of $4\times 10^{-5}$ T) and an important point is that
it never vanishes.

We assume that the field is weak, below $10^{-3}$ T, so that the
hyperfine structure remains coupled: the eigenstates are the
$\left|F,M_F \right>$ sublevels and it is a good approximation to
consider only first order Zeeman effect. As the field never
vanishes, the adiabatic theorem can be applied and the projection
$M_F$ of the angular momentum remains constant on a quantization
axis which follows the local direction of the field. The magnetic
phases $\phi(F, M_F)$ are given by:

\begin{equation}
\label{e10} \phi(M_F)  = \frac{g_F \mu_B M_F}{\hbar v} \int B(s)
ds
\end{equation}
\noindent where $g_F$ is the hyperfine Land\'e factor, $B$ is the
modulus of the magnetic field and the integral is carried along
the atomic path. Neglecting the nuclear spin contribution to the
atomic magnetic moment, for lithium $^7$Li, the nuclear spin is
$I= 3/2$ and the hyperfine levels with $F=1$ and $2$ have opposite
Land\'e factors equal to $g_F = -1/2$ for the $F= 1$ and $g_F =
+1/2$ for the $F=2$.

The magnetic phases are quite large, $\phi(M_F)/M_F = 2\times10^3$
rad for a field $B= 4 \times10^{-5}$ T. Fortunately, these phases
play no role in the absence of non-adiabatic transitions from one
sublevel to another one. The interferometer signal is only
sensitive to the phase difference for each sublevel between the
two atomic paths. In the presence of a gradient of the magnetic
field modulus $B$ in the $x$ direction, the interference pattern
corresponding to the $M_F$ level suffers a phase shift $\Delta
\phi(F,M_F) = \varphi M_F$ with $\varphi$ given by:

\begin{equation}
\label{e11} \varphi = \frac{g_F \mu_B}{\hbar v} \int
\frac{dB(s)}{dx} \Delta x(s) ds
\end{equation}
\noindent where $\Delta x(s)$ is the distance between the two
atomic paths. Let us consider a magnetic dipole $\mu$ parallel to
the ${\mathbf x}$ axis, located at a distance $d$ from the atomic
paths. We can get a closed form expression of $\varphi$ if we
neglect the homogeneous background field and if we assume that
$\Delta x(s)$ is almost constant over the region where the
gradient of $B$ is large, we get:
\begin{equation}
\label{e12} \int \frac{dB(s)}{dx} ds = \frac{\mu_0 \mu}{2 \pi d^3}
\int_{-\pi/2}^{\pi/2}\left[3\cos^2\theta +1\right]^{1/2}
\cos\theta d\theta
\end{equation}
\noindent where the integral over $\theta$ is equal to $3.42$. One
must not forget that the approximations made are not very good.
With $^7$Li hyperfine level structure, in the absence of optical
pumping, i.e. assuming the same population for the $8$ sublevels,
the interference visibility varies with $\varphi$ in the following
way:
\begin{equation}
\label{e13} {\mathcal{V}} = {\mathcal{V}}_0 \frac{2 + 4
\cos\varphi + 2 \cos 2\varphi}{8}
\end{equation}
\noindent With our approximations, $\varphi$ is a linear function
of the dipole moment or of the current if we use a coil. Moreover
$\varphi$ is proportional to $v^{-2}$, where $v$ is the atom
velocity: a $v^{-1}$ factor is obvious in equation (\ref{e11}) and
the other factor is hidden in the quantity $\Delta x(s)$
proportional to the diffraction angle.

The velocity distribution of the lithium atoms induces a
dispersion on $\varphi$ which further induces a fringe visibility
reduction. Assuming a Gaussian velocity distribution profile $P(v)
dv \propto \exp[-(v-u)^2/\alpha^2]$, we deduce a phase
distribution:
\begin{equation}
\label{e14} P(\varphi) d\varphi \propto \exp[-(\varphi-
\varphi_m)^2/\beta^2]
\end{equation}

\noindent with the phase $\varphi_m$ corresponding to the velocity
$u$ and $\beta = 2\varphi_m \alpha/u$. This approximate formula is
valid in the limit $\alpha \ll u$. After averaging over $\varphi$,
the visibility ${\mathcal{V}}$ is then still given by equation
(\ref{e13}), where $\cos (k\varphi)$ ($k$ is an integer) is
replaced by its average $\left<\cos (k\varphi\right)>$ over the
distribution $P(\varphi)$ simply given by:

\begin{equation}
\label{e15} \left<\cos (k\varphi\right)> =\cos (k\varphi_m) \exp[
-k^2\beta^2/4]
\end{equation}

\begin{figure}[hbtp]
  \begin{center}
  \includegraphics[width= 11.5 cm]{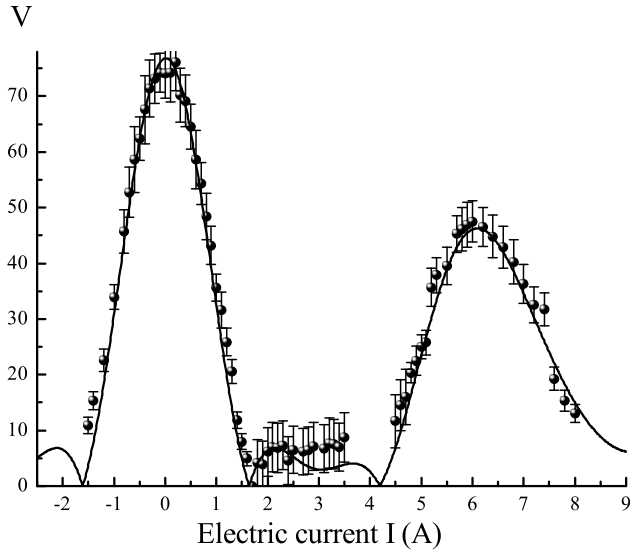}
\caption{\label{magneticgradient} Fringe visibility ${\mathcal V}$
as a function of the electric current $I$ in the coil which
creates a magnetic gradient over the atomic paths. The points are
experimental while the full curve is the best fit using equations
(\ref{e13}) and (\ref{e15}) with $\alpha/u= 0.111 $, a value
rather close but lower than the one measured on our incident
lithium beam $\alpha/u= 0.133$, showing the selectivity of the
Bragg diffraction process. The dashed curve is the curve predicted
if the velocity was perfectly defined and equal to $u$ (i.e. if
$\alpha=0$) }
  \end{center}
  \end{figure}

We have done a first experiment with a coil outside the vacuum
tank. The coil, with $350$ turns and a mean turn area close to
$50$ cm$^2$, is located at about $20$ cm from the atomic paths. We
have recorded interference fringes for different currents $I$,
varying from $0$ to $8$ A by $0.1$ A steps. We have measured the
fringe visibility ${\mathcal{V}}$, which is plotted as a function
of the current $I$ on figure \ref{magneticgradient}. Because of
the dispersion on $\varphi$ due to the velocity distribution of
the lithium atoms, the visibility observed at the peak of the
revival is not as large as when $\varphi=0$. As a consequence, the
variation of the visibility with the applied field gradient
contains an information on the velocity dispersion of the atoms
contributing to the atomic interference signal. As Bragg
diffraction is velocity selective, this velocity distribution may
differ from the velocity distribution of the lithium beam measured
at the entrance of the interferometer \cite{miffre04a,miffre04b}.
The present arrangement with a large coil rather far from the
atomic path is not very favorable for a precise analysis, because
the applied field is perturbed by the magnetic parts of the setup,
but with an improved arrangement, we hope to measure accurately
the velocity distribution of the atoms contributing to the
interference signal.

Recently, we have used a small coil under vacuum ($3.5$ turns of
wire on a $3$ cm diameter ring, with the coil center at a distance
$d = 7.5$ mm from the atomic paths). In a first time, we have
studied with care the region of near zero field gradient and we
have observed an improved fringe visibility for a small current in
the coil, thus proving that a small but non negligible magnetic
field gradient is present in our apparatus. In such an experiment,
we do not cancel everywhere the magnetic field gradient but we
simply cancel the integral appearing in equation (\ref{e11}). In
this experiment, the best observed visibility is ${\mathcal{V}} =
84.5\pm 1.0$ \% for the diffraction order $p=1$ and ${\mathcal{V}}
= 54.0 \pm 1.0$ \% for the diffraction order $p=2$ and these
results are presented in figure \ref{atomicfringes}.

The effect of an electric field gradient exists also and it has
been used recently \cite{roberts04} for the compensation of phase
dispersion in an atom interferometer. The Stark effect is
quadratic in electric field and, in a $^2S_{1/2}$ state, it is,
with an excellent approximation, independent of the $F,M_F$
sublevel as a consequence of the Wigner-Eckart theorem. The
induced phase is the same for all the $F,M_F$ levels and this
phase will play a role only if it is large, because of its
dispersion with the atom velocity. A large phase will exist only
if the electric field and its gradient are both large enough. The
stray electric field normally encountered inside vacuum chambers
are usually weak, below $1$ V/cm, and we do not expect a large
gradient, especially close to the metallic rail supporting the
mirrors. The resulting loss of coherence due to the stray electric
field should be fully negligible.

\section {Conclusion}

In this paper, we have described our Mach-Zehnder atom
inteferometer operating with a thermal lithium beam and we have
shown some examples of the observed signals. We have briefly
recalled the main theoretical points, as they are very important
to choose the best parameters. We have then given a description of
this interferometer and its operation: vacuum system, laser system
and laser standing waves, alignment procedures, the other parts
being the subjects of separate publications.

In our interferometer like in the metastable neon interferometer
of Siu Au Lee and coworkers \cite{giltner95b}, the mirrors and
beam-splitters for the atomic waves are based on elastic
diffraction by laser standing waves, in the Bragg regime. This
choice provides an almost ideal interferometer and, in agreement
with the theory of such interferometers, we have measured a high
transmission and an excellent visibility, ${\mathcal{V}} = 84.5
\pm 1.0$\%, when using first order diffraction: this is the best
visibility ever achieved with a thermal atom interferometer with
spatially separated atomic paths. This observation proves that the
atom propagation is almost perfectly coherent in the
interferometer: the two atomic paths are separated by $100$
$\mu$m, in the vicinity of the second laser standing wave and this
distance is close to $2\times 10^6$ de Broglie wavelengths !

We have also operated our interferometer with the diffraction
orders $p=2$ and $3$. The fringe visibility diminishes rapidly
with the diffraction order $p$ and we are presently investigating
the origins of this rapid diminution. We have also tested the
effect of the main misalignments on the fringe visibility, with
results in excellent agreement with theory. We have studied the
signal intensity and the fringe visibility as a function of the
width of the collimation and detector slits.

This study will serve to optimize the operating conditions and to
reach the best phase sensitivity, which is a very important point
for the accurate measurement of perturbations. We have achieved a
phase sensitivity close to $25$ mrad$/\sqrt{Hz}$ which is better
than the $34$ mrad$/\sqrt{Hz}$ obtained in our previous study
\cite{delhuille02a} (an error was made in this paper and we gave a
value which was too small by a factor $2$). With minor
improvements, we hope to measure phase shifts with an accuracy
close to $1$ mrad in a few minutes of experiment.

Finally, following previous experiments, we have applied a
magnetic field gradient: when the gradient increases, the fringe
visibility first decreases and vanishes, before presenting a
revival for a larger gradient. The intensity of the visibility
revival is a sensitive tool to measure the velocity spread of the
atoms contributing to the interferometer signal.

We are going to proceed now to interferometric measurements: our
first goals are the measurements of the electric polarizability of
lithium atom and of the index of refraction of permanent gases for
lithium waves. The possibility of using several diffractions
orders may reveal very interesting in this case, as the path
separation is proportional to the diffraction order.

\section{Acknowledgements}

We are very much indebted toward R. Delhuille, who was the first
to operate successfully this atom interferometer in 2001. We also
thank C. Champenois, L. Jozefowski for their important
contributions to the early phase of this work, and L. Lazar, for
her participation to the magnetic rephasing experiment. Special
thanks to F. Biraben and F. Nez for their loan of material,  to D.
Pritchard and Siu Au Lee for various information concerning their
experiments. We thank CNRS SPM, R\'egion Midi-Pyr\'en\'ees,
universit\'e Paul Sabatier and IRSAMC for financial support.



\end{document}